\documentclass{aastex63}

\usepackage{hyperref,epsfig,graphicx}

\received{***}
\revised{***}
\accepted{***}
\submitjournal{ApJ}
\shorttitle{Unambiguous Filament Splitting}
\shortauthors{Sun et al.}
\graphicspath{{./}{figures/}}

\begin{document}

\title{Observation of two splitting processes in a partial filament eruption on the sun: the role of breakout reconnection}

\author[0000-0001-5657-7587]{Zheng Sun}
\affiliation{School of Earth and Space Sciences, Peking University, Beijing, 100871, People's Republic of China; \href{huitian@pku.edu.cn}{huitian@pku.edu.cn}}

\author[0000-0001-6655-1743]{Ting Li}
\affiliation{National Astronomical Observatories, Chinese Academy of Sciences, Beijing 100101, People's Republic of China; \href{liting@nao.cas.cn}{liting@nao.cas.cn}}
\affiliation{School of Astronomy and Space Science, University of Chinese Academy of Sciences, Beijing 100049, People's Republic of China}

\author[0000-0002-1369-1758]{Hui Tian}
\affiliation{School of Earth and Space Sciences, Peking University, Beijing, 100871, People's Republic of China; \href{huitian@pku.edu.cn}{huitian@pku.edu.cn}}
\affiliation{National Space Science Center, Chinese Academy of Sciences, Beijing 100190, People's Republic of China}
\affiliation{National Astronomical Observatories, Chinese Academy of Sciences, Beijing 100101, People's Republic of China; \href{liting@nao.cas.cn}{liting@nao.cas.cn}}

\author[0000-0002-9534-1638]{Yijun Hou}
\affiliation{National Astronomical Observatories, Chinese Academy of Sciences, Beijing 100101, People's Republic of China; \href{liting@nao.cas.cn}{liting@nao.cas.cn}}
\affiliation{School of Astronomy and Space Science, University of Chinese Academy of Sciences, Beijing 100049, People's Republic of China}

\author[0000-0003-4804-5673]{Zhenyong Hou}
\affiliation{School of Earth and Space Sciences, Peking University, Beijing, 100871, People's Republic of China; \href{huitian@pku.edu.cn}{huitian@pku.edu.cn}}
\affiliation{Key Laboratory of Modern Astronomy and Astrophysics (Nanjing University), Ministry of Education, Nanjing 210093, People's Republic of China}

\author[0000-0001-7866-4358]{Hechao Chen}
\affiliation{School of Physics and Astronomy, Yunnan University, Kunming 650500, People's Republic of China}

\author[0000-0003-2686-9153]{Xianyong Bai}
\affiliation{National Astronomical Observatories, Chinese Academy of Sciences, Beijing 100101, People's Republic of China; \href{liting@nao.cas.cn}{liting@nao.cas.cn}}

\author{Yuanyong Deng}
\affiliation{National Astronomical Observatories, Chinese Academy of Sciences, Beijing 100101, People's Republic of China; \href{liting@nao.cas.cn}{liting@nao.cas.cn}}

\begin{abstract}
  Partial filament eruptions have often been observed, however, the physical mechanisms that lead to filament splitting are not yet fully understood. In this study, we present a unique event of a partial filament eruption that undergoes two distinct splitting processes. The first process involves vertical splitting and is accompanied by brightenings inside the filament, which may result from internal magentic reconnection within the filament. Following the first splitting process, the filament is separated into an upper part and a lower part. Subsequently, the upper part undergoes a second splitting, which is accompanied by a coronal blowout jet. An extrapolation of the coronal magnetic field reveals a hyperbolic flux tube structure above the filament, indicating the occurrence of breakout reconnection that reduces the constraning field above. Consequently, the filament is lifted up, but at a nonuniform speed. The high-speed part reaches the breakout current sheet to generate the blowout jet, while the low-speed part falls back to the solar surface, resulting in the second splitting. In addition, continuous brightenings are observed along the flare ribbons, suggesting the occurrence of slipping reconnection process. This study presents, for the first time, the unambiguous observation of a two-stage filament splitting process, advancing our understanding of the complex dynamics of solar eruptions.
\end{abstract}


\section{INTRODUCTION} \label{sec:intro}

Solar filaments are structures of cold and dense plasma suspended in the hot and tenuous solar corona \citep{devore2005solar,mackay2010physics,parenti2014solar}. With typical temperatures of around $\sim$$6000 \ K$ and densities of $\sim$$10^{11} \ cm^{-3}$ \citep{1979SoPh...61..201M,2003ApJ...586..562G,2010SSRv..151..333M}, they are usually located above magnetic polarity-inversion lines (PILs) and are believed to be supported by the magnetic tension force provided by concave-upward magnetic dips \citep{1995ApJ...443..818L,2010ApJ...718.1388D,ichimoto2023structure}. When disturbed, they can undergo eruptions and result in the formation of flares and coronal mass ejections (CMEs), which can have significant impacts on the earth and planetary space environments \citep{yan2013case,sun2022cross,yang2023sympathetic}. Therefore, investigating the eruption processes of solar filaments is essential for a better understanding of their effects on space weather.

Filament eruptions are caused by the loss of equilibrium \citep{jing2004relation,forbes2006cme,chen2011coronal} and can be triggered by two categories of physical processes: magnetic reconnection and ideal magnetohydrodynamics (MHD) instabilities. Magnetic reconnection models include flux emergence \citep{chen2000emerging,lin2001prominence,xu2008statistical}, tether-cutting \citep{moore2001onset}, breakout-type \citep{antiochos1999model,wyper2017universal,wyper2018breakout}, and catastrophic \citep{lin2000effects,lin2004cme,xie2017numerical} models. In the tether-cutting model, the reconnection point is located below the filament or magnetic flux rope (MFR), providing upward magnetic pressure \citep{sterling2003tether,chen2018witnessing}. In contrast, in the breakout-type models, the reconnection point is situated above, reducing the constraining force from above \citep{devore2008homologous,shen2012sympathetic,kumar2018evidence}. Ideal MHD models consist of kink instability and torus instability models. Kink instability occurs when the magnetic flux rope is twisted beyond a threshhold \citep{hood1981critical,fan2005coronal} and exhibits writhing motions along its axis, resulting in its ejection \citep{rust2005observational,williams2005eruption,wyper2016three}. After the filament eruption is triggered, the success of eruption often depends on the overlying magnetic field \citep{li2020magnetic,li2021magnetic}. If the magnetic restraining force is too strong, an eruption could be a failed one \citep{ji2003observations,torok2005confined,peng2022mechanism}. 

Partial filament eruptions can sometimes occur with part of the filament materials erupting outward and the other part falling back to the solar surface \citep{gilbert2000active}, and such phenomena have been simulated numerically \citep{gibson2006partial,gilbert2007filament,kliem2014slow}. 
\cite{gibson2006partial} proposed a model for the vertical splitting of filament-hosting MFRs, which involves the existence of bald patches (BPs) where magnetic field lines are tangential to the photosphere tying the MFR. The BPs exert a constraining force that prevents the lower part of the MFR from erupting, leading to the vertical splitting of the MFR. This model has been supported by observations of \citet{tripathi2009partially} and \citet{cheng2018unambiguous}.
Another widely recognized model that explains the vertical splitting of filaments is the double-decker model proposed by \citet{liu2012slow}, which was later simulated by \citet{kliem2014slow}. This model suggests that before the eruption, the filament consists of two vertically-distributed segments. The occurrence of magnetic reconnection between the segments caused their separation, ultimately resulting in the observed vertical filament splitting. This scenario has also been supported by several observations \citep{2018A&A...619A.100H,2019ApJ...872..109A,hu2022spectroscopic}.

The magnetic reconnection that results in solar eruptions is three-dimensional (3D) intrinsically. In 3D scenario, magentic reconnection occurs at 3D null-points \citep{priest1996magnetic,priest2009three} and quasi-separatrix layers (QSLs; \citealt{priest1995three,demoulin1996quasi}). QSLs are regions with large gradient of magnetic connectivity and often wrap around MFRs \citep{aulanier2005current,wilmot2009magnetic,parnell20103d}. The footprints of QSLs on the photosphere typically coincide with flare ribbons and often exhibit a hook-shaped structure \citep{janvier2013standard,li2018two}. QSL reconnection is often characterized by the progressive slipping of magnetic field lines, leading to a continuous brightening of the QSL footprints. This phenomenon has been substantiated through a multitude of simulations \citep{aulanier2006slip,finn2014quasi,effenberger2016simulations,aulanier2019drifting} and observations \citep{li2016slipping,li2018three,li2018two}.

Coronal jets are transient, narrow bursts of plasma that can be observed at different wavelengths \citep{shibata1996coronal,nistico2009characteristics,moore2010dichotomy,tsiropoula2012solar,raouafi2016solar}. Flux emergence \citep{shibata1994gigantic,isobe2007ellerman,archontis2010recurrent,chen2015recurrent,joshi2020case} and flux converging models \citep{priest1994converging,schrijver1997sustaining,attie2016relationship} are often used to explain the trigering of coronal jets. Another model is the embedded-bipole model \citep{pariat2009model,pariat2010three,pariat2015model,pariat2016model,wyper2016three}, which has an initial bipolar magnetic field configuration. In this model, helicity is injected into this field through footpoint rotation, and the kink instability induces transient energy transfer from the twisted closed field to the open field, resulting in a blowout jet. However, some coronal jets are associated with mini-filaments, as observed by \cite{sterling2015small}. To account for this phenomenon, \cite{wyper2017universal,wyper2018breakout} simulated the scenario of blowout jets with mini-filaments. Such phenomena are called breakout jets because they can be regarded as a miniature version of breakout CMEs. The breakout CME model proposes that a null point exists above the MFR in the background quadrapole magnetic field. When the structure is disturbed, a breakout reconnection occurs, reducing the upper constraint and causing the MFR to explode upward \citep{antiochos1999model}. Similar to breakout CME model, the breakout jets have an anemone configuration initially and a null point above the mini-filament. When disturbed, a breakout reconnection occurs, causing the mini-filament to rise and reconnect with the open magnetic field lines to form a blowout jet. Such a scenario has been confirmed through several observations \citep{kumar2018evidence,2019ApJ...887..246D,shen2019round,zhou2021sympathetic}.

Despite a large number of observations of partial filament eruptions, the physical mechanisms that lead to filament splitting are not yet fully understood. In this study, we present a partial filament eruption event with two splitting processes and provide a detailed analysis of their trigering and eruption mechanisms. The remainder of this paper is organized as follows. Section \ref{sec:observations} describes the observations and data analysis used in our study. In Section \ref{sec:results}, we present our observational results in detail. Finally, we discuss and summarize the major results in Section \ref{sec:discussion}.

\section{OBSERVATIONS AND DATA ANALYSIS} \label{sec:observations}
The Atmospheric Imaging Assembly (AIA) onboard the Solar Dynamics Observatory (SDO) has been providing both extreme ultraviolet (EUV) and ultraviolet (UV) images of the Sun since 2010, capturing atmospheric dynamics in a wide range of temperatures from $\lg (T/$K$) \approx 4.0$ to $\lg (T/$K$) \approx 7.0$ \citep{lemen2012atmospheric}. We analyzed 131~\AA, 171~\AA, 193~\AA, 211~\AA, and 304~\AA\ data with a cadence of 0.6$^{\prime\prime}$ pixel $^{-1}$ and a temporal resolution of 12 seconds in our study. 
In addition, the Helioseismic and Magnetic Imager (HMI) onboard SDO can capture line-of-sight(LOS) magnetograms of the entire solar disk with a spatial resolution of 0.5$^{\prime\prime}$ pixel $^{-1}$ and a cadence of 45 seconds \citep{scherrer2012helioseismic}.
We also used observations from the Solar Upper Transition Region Imager (SUTRI) and the Chinese H$\alpha$ Solar Explorer (CHASE) to analyze the dynamic evolution of an erupting filament and the accompanied blowout jet.
SUTRI is a space-based EUV imager onboard the SATech-01 satellite launched by China in 2022 \citep{2023RAA....23f5014B}. SUTRI uses the Ne VII 46.5 nm spectral line, which forms at a temperature of $\sim$0.5 $MK$, to image the upper transition region of the sun \citep{tian2017probing}. With a field of view of 41.6$^{\prime}$ $\times$  41.6$^{\prime}$, SUTRI has a spatial resolution of $\sim$8$^{\prime\prime}$ and a time resolution of $\sim$30 seconds. As the satellite is not a solar-dedicated one, its observation time is not continuous but rather limited to approximately two-thirds of each orbit (60 minutes out of a total 96 minutes). Thanks to its unique spectral band, SUTRI observations may be used to establish a link between the lower solar atmosphere and corona.
CHASE is a space-based satellite that was launched in 2021 and is designed for H$\alpha$ spectral imaging \citep{li2022chinese}. Onboard the CHASE, the H$\alpha$ Imaging Spectrograph (HIS) enables spectral imaging of both H$\alpha$ at 6559.7~\AA\ - 6565.9~\AA\ and Fe I at 6567.8~\AA\ - 6570.6~\AA. The instrument offers two scanning modes, namely Raster Scanning Mode (RSM) and Continuum Imaging Mode (CIM). For the H$\alpha$ observation in this study, the RSM mode was employed. The instrument features a spectral line with a half-width of 0.072 Å, a pixel spectral resolution of 0.024 Å, and a pixel spatial resolution of 0.52$^{\prime\prime}$.

To obtain the 3D coronal magnetic field, we utilized the photospheric vector magnetic field provided by the Space-Weather HMI Active Region Patches (SHARP; \citealt{bobra2014helioseismic}) and applied a non-linear force-free field (NLFFF) extrapolation with it. The NLFFF extrapolation was carried out by using an optimization approach, originally proposed by \citet{wheatland2000optimization}, and subsequently improved by \citet{wiegelmann2004optimization} and \citet{wiegelmann2012should}, which allows us to reconstruct the 3D coronal magnetic field. To ensure that our input data satisfies the force-free assumption, we used the preprocessing method introduced by \citet{wiegelmann2006preprocessing}.
The extrapolation box was set at 964 ${\times}$ 648 ${\times}$ 256, with a grid resolution of 0.5$^{\prime\prime}$. Using the extrapolated field, we calculated the squashing factor Q and twist number $T_w$ using the procedure developed by \citet{liu2016structure}. Q value describes the magnetic connectivity of the coronal magnetic field \citep{priest1995three}. Regions with high-Q value are referred to as QSLs\citep{demoulin1996quasi,titov2007generalized}. $T_w$ reflects the degree of twisting of the magnetic field lines, which can help us to identify the MFR \citep{berger2006writhe}.

\begin{figure}
  \centering
    \plotone{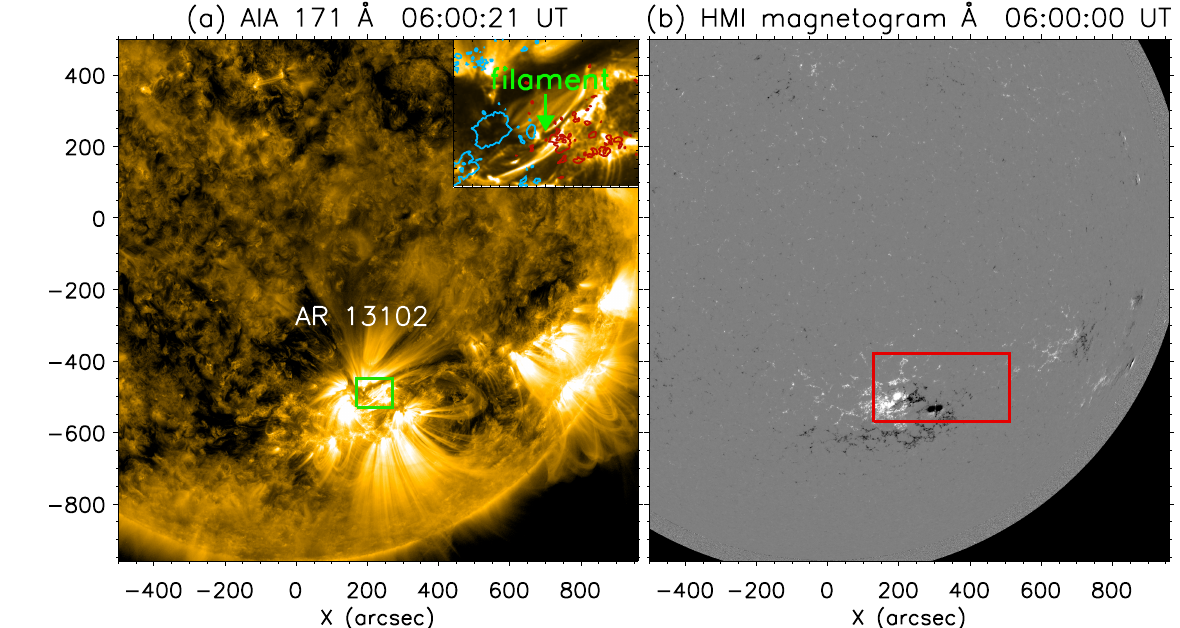}
    \caption{The overview of AR 13102 including observations of AIA 171~\AA\ (a) and HMI line-of-sight (LOS) magnetic field (b). The green box in panel (a) indicates the location of the zoom-in area in the upper right corner and also denotes the field of view (FOV) of Figure \ref{fig:Fig.3}. The zoom-in region in panel (a) was overlaid with contours of the magnetic field at $\pm$350 G levels. The red box in panel (b) denotes the FOV of Figure \ref{fig:Fig.2}
  \label{fig:Fig.1}}
\end{figure}

\begin{figure}
  \centering
  \plotone{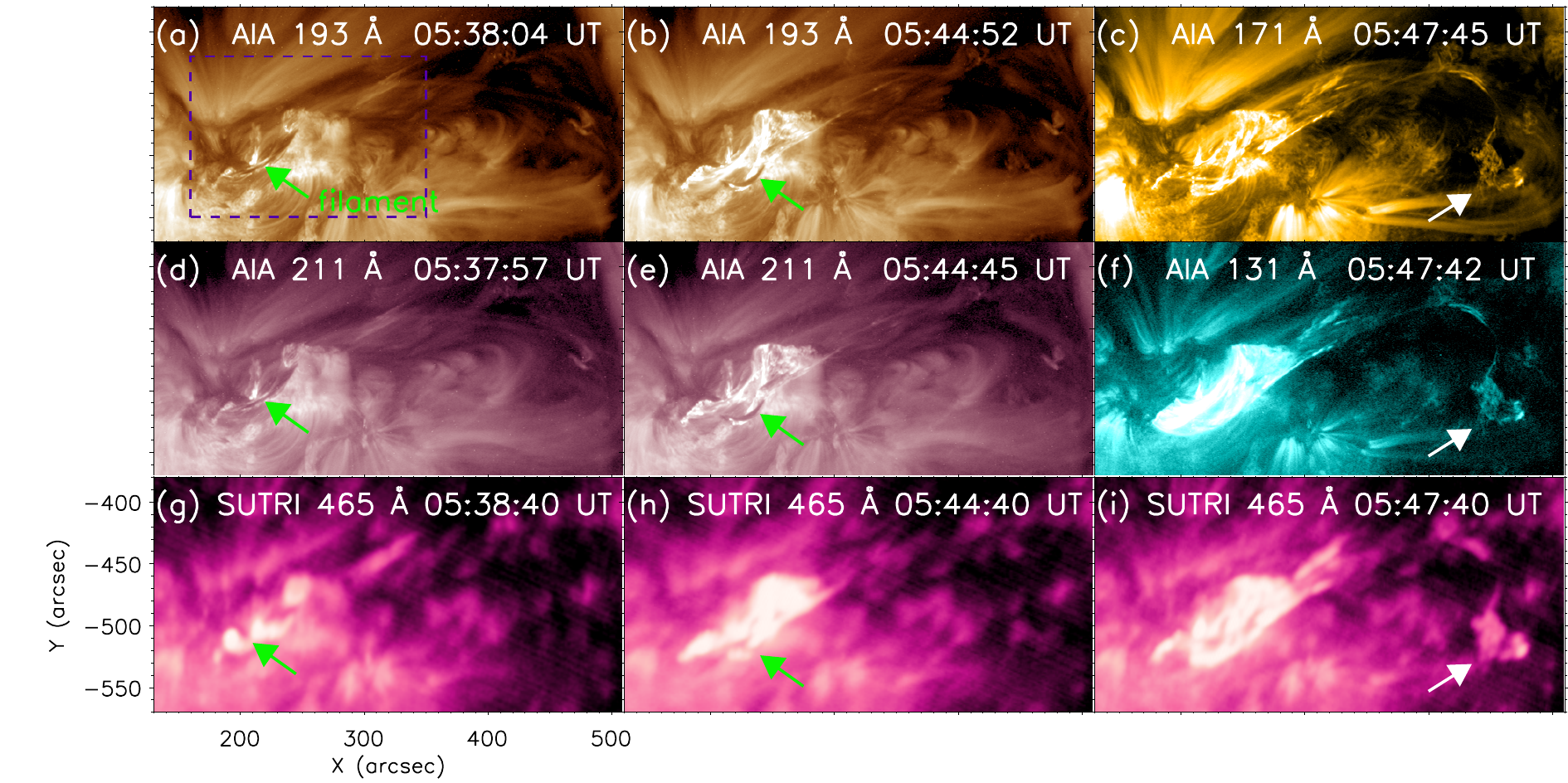}
  \caption{The entire process of the filament eruption observed at various wavelengths. The purple box in panel (a) denotes the FOV of Figure \ref{fig:Fig.4}. The green arrows in the left and middle columns point to the filament. The white arrows in the right column point to the remote brightening. (An animation from 05:30 UT to 06:19 UT of this figure is available online, which contains the entire process of the filament eruption in 211~\AA, 193~\AA, 171~\AA, and 131~\AA\ channels.)
  \label{fig:Fig.2}}
\end{figure}

\begin{figure}
  \centering
  \plotone{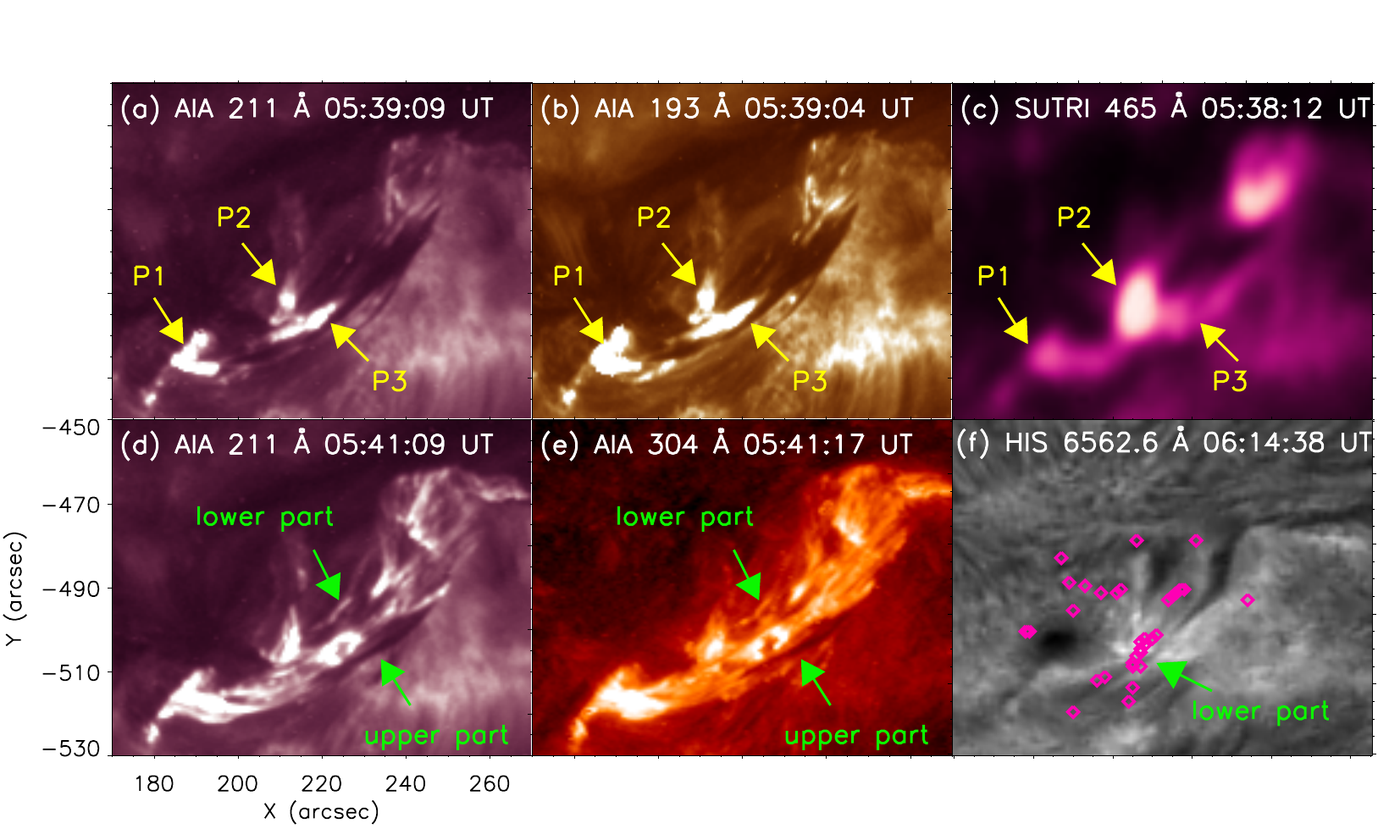}
  \caption{The first splitting process of the filament observed by SDO/AIA, SUTRI and HIS. Panels (a)-(c): brightenings associated with the initial rise and the first splitting of the filament . The yellow arrows denote the location of the brightenings. Panels (d)-(e): the filament is separated into the lower part and the upper part. Panel (f): post-eruption H$\alpha$ observation showing that the lower part remained in place. The pink diamonds in panel (f) denote the locations of bald patches.
  \label{fig:Fig.3}}
\end{figure}

\begin{figure}
  \centering
  \plotone{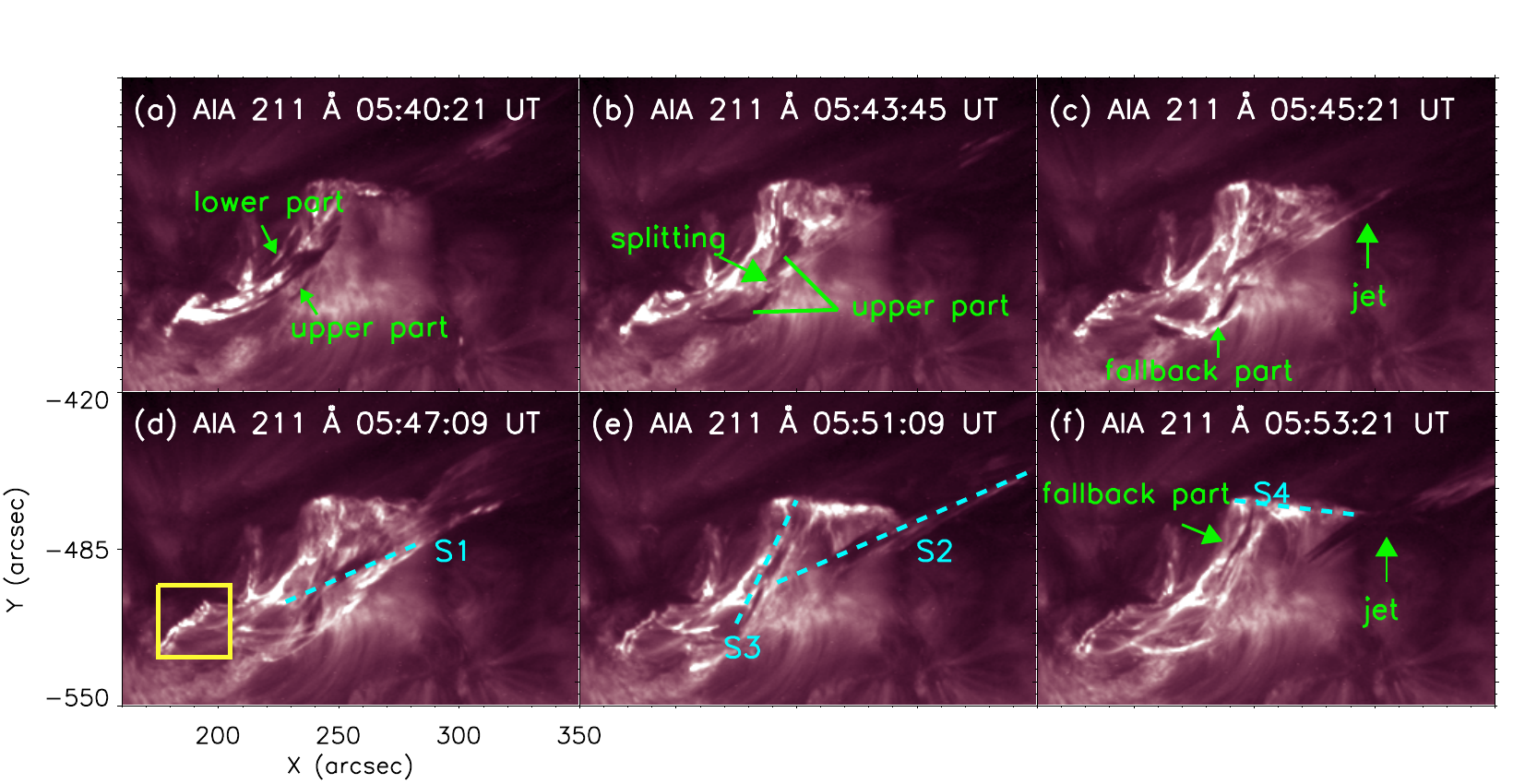}
  \caption{The second splitting process of the filament accompanied by a blowout jet. The yellow square in panel (d) denotes the FOV of Figure \ref{fig:Fig.6}. The slices S1-S4 in panels (d)-(f) are chosen to make time-distance diagrams shown in Figure \ref{fig:Fig.5}. S1-S4 are chosen to show, respectively, the splitting process of the filament, the eruption of the coronal jet, the trajectory of the fallback part and the continous brightening along the northern flare ribbon. (An animation from 05:30 UT to 06:06 UT of this figure is available online, which presents the second splitting process of the filament in 211~\AA.)
  \label{fig:Fig.4}}
\end{figure}

\begin{figure}
  \centering
  \plotone{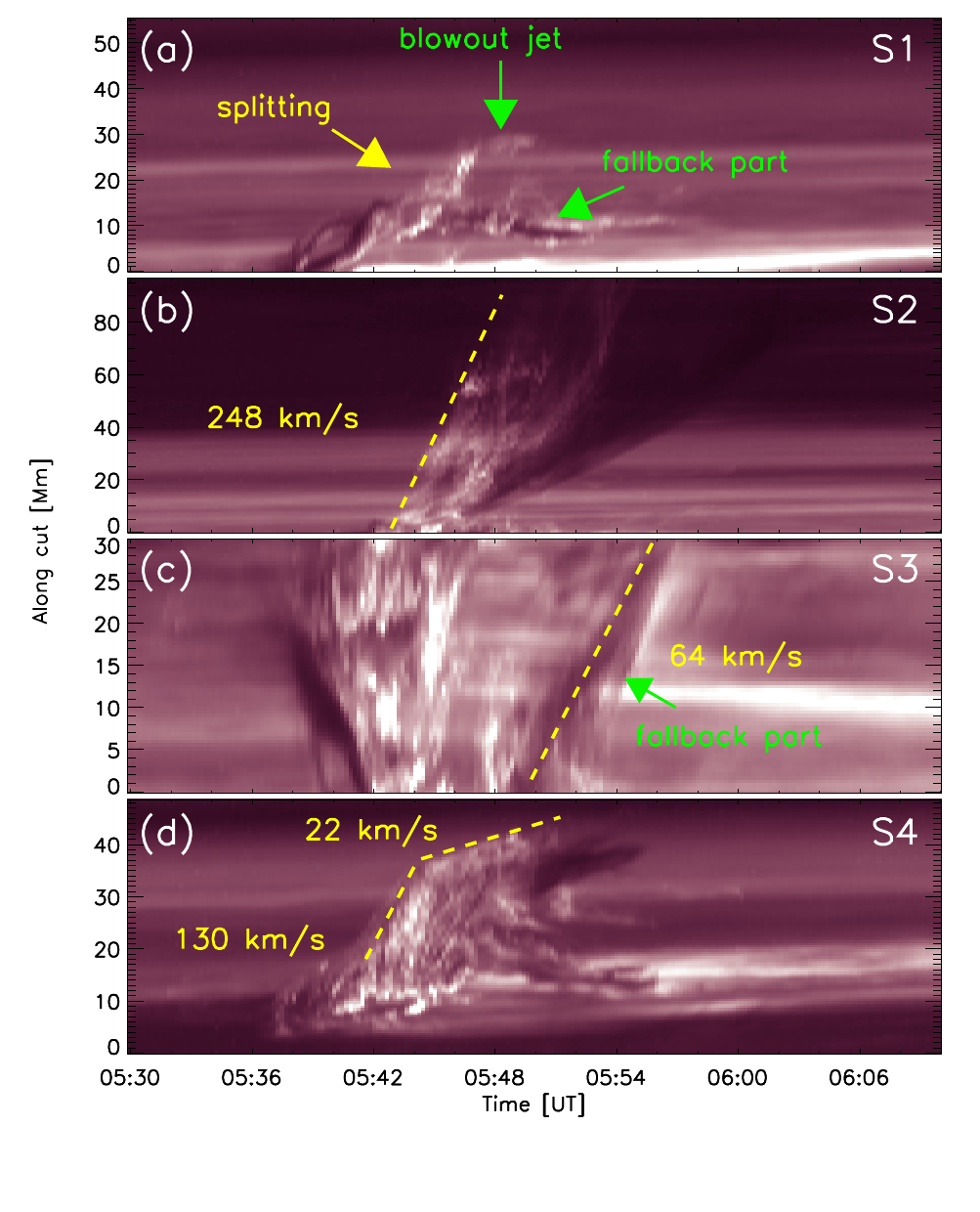}
  \caption{Time-distance diagrams of AIA 211~\AA\ images for the slices S1-S4 in Figure \ref{fig:Fig.4}. The yellow dashed lines are plotted as a result of a linear fit.
  \label{fig:Fig.5}}
\end{figure}

\begin{figure}
  \centering
  \plotone{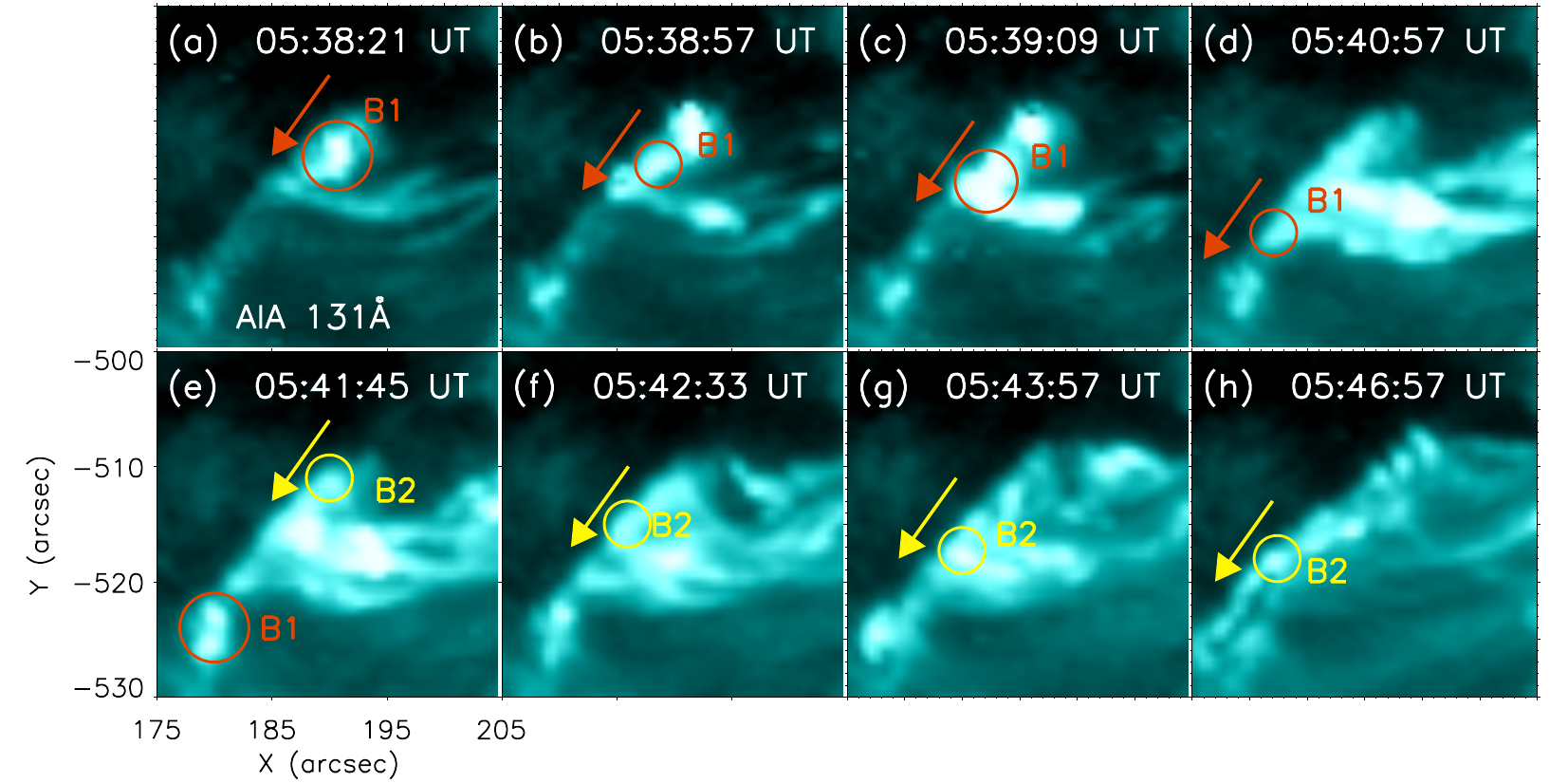}
  \caption{The successive brightenings along the southern flare ribbon. The red and yellow circles indicate the location of the brightenings and the arrows denote the direction of their motions. (An animation from 05:30 UT to 05:59 UT of this figure is available online, which shows the successive brightenings along the southern flare ribbon in 131~\AA.)
  \label{fig:Fig.6}}
\end{figure}

\begin{figure}
  \centering
  \plotone{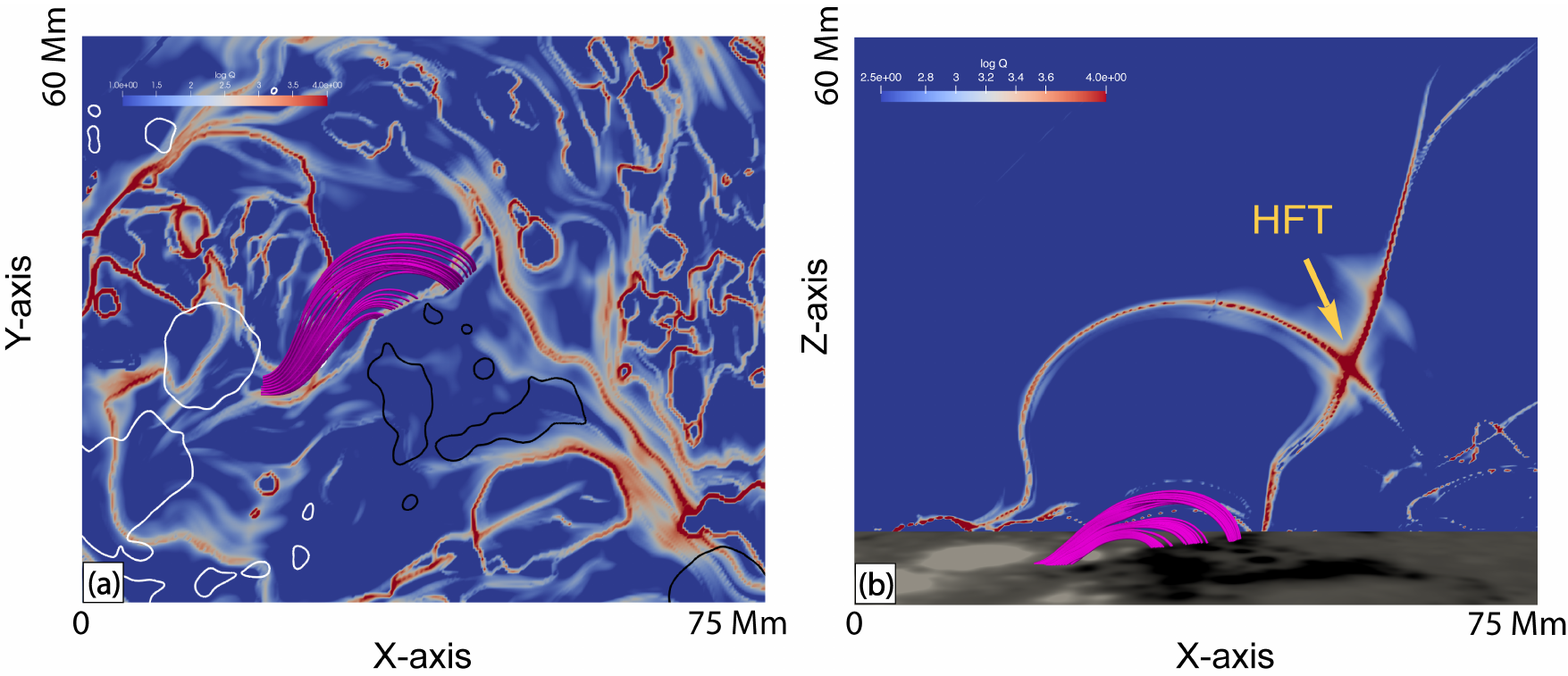}
  \caption{Q-value maps of the AR 13102 and the extrapolated structure of MFR. Panel(a): distribution of logarithm Q on the bottom boundary. The pink tubes indicate the MFR. White and Black contours represent the magnetic field at $\pm$500 G levels. Panel(b): distribution of logarithm Q on the x-z plane (y=40 Mm in panel (a)). The bottom boundary is the LOS magnetogram. The orange arrow denotes the location of the hyperbolic flux tube (HFT).
  \label{fig:Fig.7}}
\end{figure}

\section{RESULTS} \label{sec:results}
On 2022 September 20, NOAA active region (AR) 13102 is located at S26$^\circ$W27$^\circ$ on the southwest part of the solar disk. Figure \ref{fig:Fig.1} provides an overview of this active region, with the zoom-in region of panel (a) highlighting a distinct filament. By comparing the magnetogram presented in panel (b), we can see that the filament is situated above the PIL. 

An overview of the filament eruption process is presented in Figure \ref{fig:Fig.2}. The eruption is accompanied by a C5.5-class flare that begins at $\sim$05:35 UT and peaks at $\sim$05:42 UT. The filament begins to ascend at $\sim$05:38 UT and undergoes the first splitting (panels (a), (d), (g) and Figure \ref{fig:Fig.3}). At $\sim$05:44 UT, the rising filament begins to split for the second time, as clearly evidenced by observations at 193~\AA\ and 211~\AA, as well as the associated online animation. The continued ascent of the filament results in the formation of a coronal jet, as shown in the middle and right columns in Figure \ref{fig:Fig.2}. A remote brightening is visible along the trajectory of the jet, marked by a white arrow in panels (c), (f), and (i). Previous investigations have found that such remote brightenings are associated with the spine of a magnetic null point connecting (e.g., \citealt{sun2013hot,li2018two,song2018investigation}). The observations indicate that the jet is likely to be classified as a breakout jet, which involves breakout magnetic reconnection between the filament and the ambient field \citep{wyper2017universal,wyper2018breakout}.

Associated with the the first splitting process of the filament, several brightenings can be observed, as shown in Figure \ref{fig:Fig.3} (a), (b), and (c). P1-P3 brightenings are detected at 211~\AA, 193~\AA, and 465~\AA, with P1 located at the eastern footpoint of the filament and P2-P3 at the main body of the filament.
Starting at $\sim$05:41 UT, the filament can be observed to split into two parts, as clearly seen in panels (d) and (e), with the upper part being ejected upwards while the lower part appeares to remain in place. The H$\alpha$ data from CHASE is available only after the eruption, as shown in panel (f). It is revealed that a part of the filament remains in the AR, which corresponds to the lower part of the filament that is not ejected away. BPs are the regions with strong constraining between the lower part of the filament and solar surface. Using the definition, $(\mathbf{B} \cdot \nabla) B_z>0$ and $B_z=0$, given in \citet{titov1993conditions} and the magnetic field vector provided by SHARP, we calculated the locations of the BPs (pink diamonds in panel (f)). The concentrated distribution of BPs below the filament demonstates that the lower part is bounded tightly \citep{gibson2006partial}. Moreover, the splitting location is found to be exactly at the location of the P3 brightening. This suggests that P3 is caused by magnetic reconnection inside the filament, resulting in a vertical splitting, similar to the finding of \citet{cheng2018unambiguous}. 

Following the first splitting process, the filament is separated into an upper and a lower part. Subsequently, the upper part undergoes a second splitting as depicted in Figure \ref{fig:Fig.4}. Prior to the splitting of the upper part, some faint outflows can be observed in the northwest of the filament (see Figure \ref{fig:Fig.4}(a) and the associated animation). These pre-eruption outflows are similar to the observations in \citet{2021ApJ...907...41K}, which suggested that these pre-eruption outflows are indicative of the occurrence of breakout reconnection At $\sim$05:43 UT, a distinct splitting point, denoted by the green arrow in panel (b), is observed. An examination of panels (c)-(f) and the accompanying online animation reveals that the southern segment of the filament is elevated to a certain altitude and then falls back to the solar surface. This segment is referred to as the fallback part. In contrast, the northern segment continues to rise, eventually leading to a blowout jet.
To examine the splitting process of the upper part of the filament, a time-distance diagram (Figure \ref{fig:Fig.5}(a)) was generated along slice S1, marked in Figure \ref{fig:Fig.4}(d). This diagram clearly illustrates the splitting of the upper part of the filament. Specifically, while one part of the filament ascends to form the coronal jet, the other part falls back down. Another time-distance diagram along S2 was created to examine the dynamics of the coronal jet, as illustrated in Figure \ref{fig:Fig.5}(b). The leading edge of the jet moves at a speed of $\sim$248 km s$^{-1}$. The animation further reveals that the initial formation of the coronal jet involved a distinct untwisting helical process, indicating that the magnetic helicity of the filament is transported into the upper atmosphere.  The stack plot in Figure \ref{fig:Fig.5}(b) also indicates that the hot component of the jet is faster than the cool component, consistent with the finding of \citet{joshi2017observational}. Additionally, to study the fallback part's kinematics, slice S3 was chosen along its path, as shown in Figure \ref{fig:Fig.5}(c). A linear fit of the trajectory of the fallback part yields a projected speed of $\sim$64 km s$^{-1}$.

Beginning at $\sim$05:38 UT, observations reveal the formation of flare ribbons, indicating the heating caused by energetic electrons produced by the magnetic reconnection above. The animation reveals successive brightening along flare ribbons. For further investigation of this phenomenon, slice S4, was chosen along the northern flare ribbon, and the corresponding time-distance diagram was generated presented in Figure \ref{fig:Fig.5}(d). This diagram reveals the elongation motion of the brightening, which indicates the occurrence of 3D magnetic reconnection. The apparent slipping speed is 130 km s$^{-1}$ at the initial stage and decreases to about 20 km s$^{-1}$ after $\sim$05:44 UT. We also noticed that the beginning time of the elongation of the northern flare ribbon correlated well with the time of filament's initial rise, suggesting that the breakout reconnection played a role in the initial rise of the filament. The southern flare ribbon also exhibites successive brightenings, as shown in Figure \ref{fig:Fig.6}. The motion of B1 along the red arrow indicates a successive brightening process caused by heating from energetic electrons generated by the reconnection site. Similarly, B2, marked in yellow, appeares at $\sim$05:41 UT and moves along the same direction as B1, revealing elongation motion in the southern flare ribbon and indicating the occurrence of slipping magnetic reconnection \citep{aulanier2019drifting}.

In order to investigate the 3D coronal magnetic field of this AR, a NLFFF extrapolation was performed. Based on the extrapolated 3D magnetic field, we calculated the squashing factor Q and twist number $T_w$. Using the $T_w$ map on the photosphere, we selected the regions where $|T_w| \ge 1$ and tracked the magnetic field lines. We saw a flux rope structure as marked in pink in Figure \ref{fig:Fig.7}(a). We also plotted the vertical slice of the Q value shown in Figure \ref{fig:Fig.7}(b). The results reveal the presence of two QSLs crossing diagonally above the flux rope. To ascertain whether this crossing constitutes a genuine null point, we employ the trilinear method proposed by \citet{haynes2007trilinear} and implemented by \citet{federica_chiti_2020_4308622} to calculate the positions of the magnetic null points. The analysis reveals that the crossing between the two QSLs does not possess a null point structure. However, this configuration of Q-value map indicates the presence of a hyperbolic flux tube (HFT) structure, which is also a magnetic structure conducive to facilitating magnetic reconnection. The remote footpoint of the HFT corresponds to the remote brightening shown in Figure \ref{fig:Fig.2}(c), (f) and (i). Such a structure may form a breakout current sheet, where the filament can reconnect with the ambient field to form the blowout jet.

\begin{figure}
  \centering
  \plotone{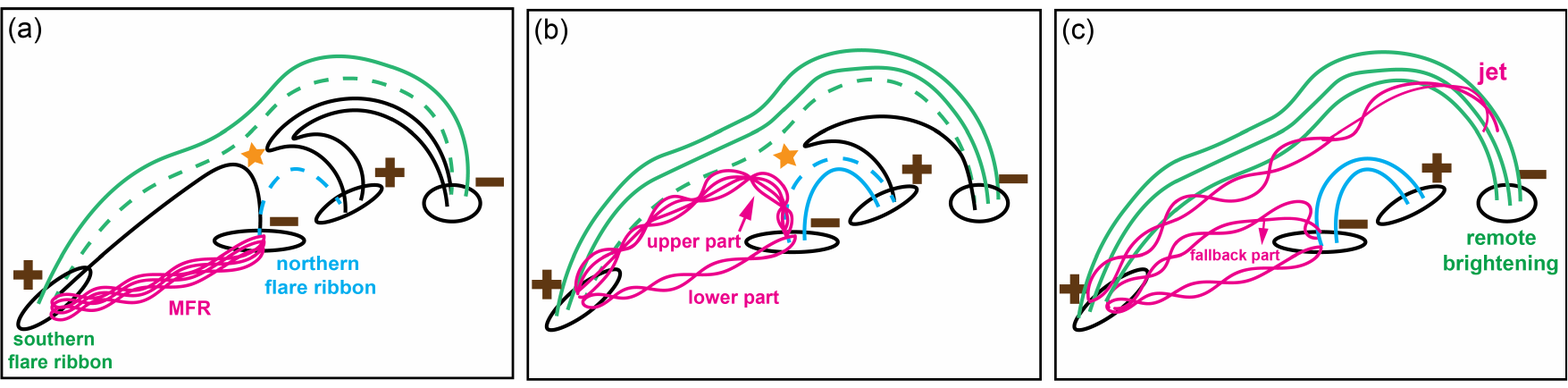}
  \caption{Schematic sketch of the eruption process. Black lines show the magnetic field lines involved in breakout reconnection and the star symbol shows the breakout reconnection site. The green and blue dashed lines represents the newly formed magnetic field lines due to the breakout reconnection. The pink lines indicate the MFR and the pink arrows point to the direction of its motion.
  \label{fig:Fig.8}}
\end{figure}

\section{SUMMARY AND DISCUSSION} \label{sec:discussion}
Our observations unambiguously show that a filament located in active region underwent two splitting processes. The first process is a vertical splitting with brightenings observed within the filament. As a result of this splitting, the filament is separated into an upper and a lower part. The upper part continues to lift, while the lower part remains in place. During the rise of the upper part, it undergoes a second splitting, which is clearly imaged. A portion of the upper part continues to rise and generates a blowout jet, while the remaining part falls back to the footpoint along the magnetic structure. In addition, continous brightenings along two flare ribbons are observed.

A magnetic structure with $|T_w| \ge 1$ can be regarded as a flux rope \citep{hou2019secondary}, so we can conclude that the observed filament is within a MFR. 
The first splitting process is a vertical splitting, consistent with the study of \citet{gibson2006partial}, which proposed that BPs tie the bottom part of the MFR to prevent its eruption and form a vertical splitting. The BPs surrounding the filament in Figure \ref{fig:Fig.3}(f) indicate close tying of the lower part of the filament, which is difficult to erupt due to magnetic confinement. H$\alpha$ observations after the eruption show that the lower part remains in place (Figure \ref{fig:Fig.3}(f)), providing compelling evidence that the lower part does not erupt. Brightening within the filament (P3 in Figure \ref{fig:Fig.3}) indicates internal magnetic reconnection, also in line with \citet{gibson2006partial}. Another explanation for vertical splitting is the double-decker configuration, where the MFR is already separated before the eruption and there is a X-type structure between the two parts of MFRs. However, we were unable to identify such a structure from our observations.

The second filament splitting occurs during the eruption of the upper part and is accompanied by a blowout jet. The HFT structure of this region is indicated by the magnetic field extrapolation and the remote brightening observed in Figure \ref{fig:Fig.2}. There exists a crossing between two QSLs above the MFR, suggesting the existence of a HFT above the MFR. The generation process of the blowout jet is consistent with the simulations of \citet{wyper2017universal,wyper2018breakout}. In this scenario, when the initial field is disturbed, the field lines beneath the HFT will expand towards the HFT, creating a breakout current sheet where the restraining field is reconnected with the open field. This removes the constraining force above, and the filament is lifted up. In addition, the pre-splitting outflows depicted in Figure \ref{fig:Fig.4}(a) further support this scenario. The filament then reconnects with the open ambient field when it is lifted to the breakout current sheet, forming an unwinding breakout jet. It should be noted that although their model assumes the presence of a null point structure, HFT reconnection exhibits similarities to null-point reconnection. Therefore, this model is also applicable in the context of HFT reconnection. To our knowledge, there has been no clear observation of a splitting filament in the breakout jet process. The splitting phenomenon can be explained by asymmetric filament eruption. From the online animation, it is evident that the western part of the filament is lifted higher with a larger velocity, allowing it to reach the breakout current sheet and generate the jet, while the eastern part of the filament with a smaller velocity can not reach the breakout sheet, and thus, falls back down to the solar surface. This difference in velocity is also evident in Figure \ref{fig:Fig.5}(a). \citet{liu2018disintegration} also reported a filament disintegration event, where the filament reconnected with the QSL dome and fell back down to the surface. The absence of the jet in their study indicated that there was no breakout current sheet, which is different from our observations.

To provide a more direct visualization of the scenario, a schematic sketch was created and presented in Figure \ref{fig:Fig.8}. Panel (a) illustrates the initial process of the breakout reconnection. The HFT structure, where breakout reconnection occurs, is positioned above the MFR. The breakout reconnection results in the decrease of the constraining force above the MFR, and thus the MFR starts to rise. However, due to the presence of BPs under the filament, the lower part of the filament is tied to solar surface. This leads to the splitting of the filament into an upper and a lower part (panel (b)), which represents the first splitting process.
Subsequently, a portion of the upper part is able to reach the breakout current sheet and reconnects with the ambient field, resulting in the formation of a blowout jet and the brightening of the remote footpoint, as shown in panel (c). Due to the nonuniform lifting speed of the upper part, the other portion of the upper part is unable to reach the breakout current sheet and instead falls back, which is called the fallback part. This represents the second splitting process.

We observed continuous brightening along the northern flare ribbon with an initial speed of 130 km s$^{-1}$ and a subsequent speed of 22 km s$^{-1}$, as shown in Figure \ref{fig:Fig.5}(d). Furthermore, in Figure \ref{fig:Fig.6}, we observed moving brightenings along the southern flare ribbon, indicating successive reconnection of the magnetic field lines. These observational signatures suggest the occurrence of the breakout and slipping reconnection process. \citet{li2018two} also observed similar phenomenon suggesting the existence of slipping process in a breakout jet. In the breakout jet scenario, there are two current sheets: the breakout sheet above the MFR and the thin sheet beneath the MFR (see Figure 3 in \citet{wyper2017universal}). It is debatable which current sheet dominates the shift of the footpoint and the slipping process. However, we found that the beginning time of the brightening shift is correlated with the beginning time of the initial rise of the filament caused by the reconnection in the breakout current sheet, as shown in Figure \ref{fig:Fig.5}(d). This correlation suggests that the breakout current sheet dominates the slipping process rather than the underlying sheet. Moreover, in \citet{wyper2017universal,wyper2018breakout}, it is convincingly demonstrated that the breakout reconnection contributes more energy than the lower reconnection, which is in agreement with our observation.

\acknowledgments
 This work is supported by the National Key R\&D Program of China (2019YFA0405000),the National Natural Science Foundations of China (11825301, 12222306, and 12273060), the B-type Strategic Priority Program of the Chinese Academy of Sciences (XDB41000000), the Youth Innovation Promotion Association of CAS (2023061 and 2023063) and Yunnan Academician Workstation of Wang Jingxiu (No. 202005AF150025). SUTRI is a collaborative project conducted by the National Astronomical Observatories of CAS, Peking University, Tongji University, Xi'an Institute of Optics and Precision Mechanics of CAS and the Innovation Academy for Microsatellites of CAS. The CHASE mission is supported by China National Space Administration (CNSA). AIA is a payload onboard \emph{SDO}, a mission of NASA's Living With a Star Program.

\bibliographystyle{aasjournal}
\bibliography{ref}

\end{document}